\documentclass[final,authoryear,3p,times,twocolumn]{elsarticle}
\usepackage{graphics}
\usepackage{graphicx}
\usepackage{epsfig}
\usepackage{amssymb}

\journal{New Astronomy}
\begin{document}
\begin{frontmatter}
\title{BVRI lightcurves of supernovae SN 2011fe in M101, SN 2012aw in M95,\\ 
and SN 2012cg in NGC 4424}

\author[um]{U. Munari}
\author[zz]{A. Henden}
\author[sd]{R. Belligoli}
\author[sd]{F. Castellani}
\author[sd]{G. Cherini}
\author[sd]{G. L. Righetti}
\author[sd]{A. Vagnozzi}

\address[um]{INAF Astronomical Observatory of Padova, 36012 Asiago (VI), Italy; ulisse.munari@oapd.inaf.it}
\address[zz]{AAVSO, 49 Bay State Road, Cambridge, MA 02138, USA}
\address[sd]{ANS Collaboration, c/o Osservatorio Astronomico, via dell'Osservatorio 8, 36012 Asiago (VI), Italy}

\begin{abstract}
Accurate and densely populated $B$$V$$R_{\rm C}$$I_{\rm C}$ lightcurves of
supernovae SN 2011fe in M101, SN 2012aw in M95 and SN 2012cg in NGC 4424 are
presented and discussed.  The SN 2011fe lightcurves span a total range of 342
days, from 17 days pre- to 325 days post-maximum.  The observations of both
SN 2012aw and SN 2012cg were stopped by solar conjunction, when the objects
were still bright.  The lightcurve for SN 2012aw covers 92 days, that of SN
2012cg spans 44 days.  Time and brightness of maxima are measured, and from
the lightcurve shapes and decline rates the absolute magnitudes are obtained,
and the derived distances are compared to that of the parent galaxies.  The
color evolution and the bolometric lightcurves are evaluated in comparison
with those of other well observed supernovae, showing no significant
deviations.  
\end{abstract}

\end{frontmatter}

\section{Introduction}
\label{}

Most supernovae are discovered at cosmological distances. Rarely do they
become as bright as SN 2011fe, a type Ia supernova that recently erupted in
M101 and peaked at B=9.9.  In recent times, particularly bright supernovae
were SN 1937c (a type Ia in IC 4182, peaking at B=8.7; Schaefer 1996), SN
1972e (type Ia in NGC 5253, peaking at B=8.6; Branch et al.  1983), SN 1993j
(type II in M81, reaching B=10.8; Lewis et al.~1994), and of course SN
1987a (a type II in LMC, peaking at B=4.6; Hamuy et al.  1988).  Such bright
SNe are subject to "all-out" observing campaigns, and the great amount of
data collected over a wide range of wavelengths make them critical tests for
models.  The quest for the ultimate set of observational data on well
observed supernovae may go on for years or decades after the event, with
increasingly finer and more sophisticated comparisons, homogenization,
re-calibration, and merging of independent sets of data (Clocchiatti et 
al.~2011 have only recently obtained what they termed the "ultimate" lightcurve
of SN 1998bw, a key supernova that together with SN 2003dh provided the
first solid evidence of the connection between supernovae and Long-Soft
gamma-ray bursts).

The aim of this paper is to present our rich set of accurate $B$$V$$R_{\rm
C}$$I_{\rm C}$ photometric measurements of SN 2011fe, covering 342 days,
augmented by similar photometry of other two recent and bright supernovae,
SN 2012aw (type IIP) and SN 2012cg (type Ia), whose monitoring was stopped
by solar conjunction while they were still in their early decline stages.

It is known (eg.\,Suntzeff\,et\,al.\,1988, Clocchiatti\,et\,al. 2011) that time
dependent differences in the lightcurves of supernovae obtained by different
authors are the results of slightly different local realizations of the
standard photometric passbands combined with the profoundly non-stellar
character of supernova spectra (dominated by time-variable strong absorption
and emission).  An enhancing feature of our set of $B$$V$$R_{\rm C}$$I_{\rm
C}$ photometric measurements is that they have been independently and in
parallel obtained with several different telescopes equipped with different
CCD cameras and $B$$V$$R_{\rm C}$$I_{\rm C}$ filter sets, all adopting the
same accurate photometric sequences calibrated by Henden et al.~(2012) on
equatorial Landolt (1983, 1992) standards.  Combining data from different
telescopes into a single, merged, densely populated lightcurve effectively
improves upon the disturbing effects of differences in the local
realizations of a common photometric system, offering a merged lightcurve
which should be a closer match to the {\it true} lightcurve than any local
realization.

  \begin{figure}[!Ht]
    \centering   
    \includegraphics[width=8cm]{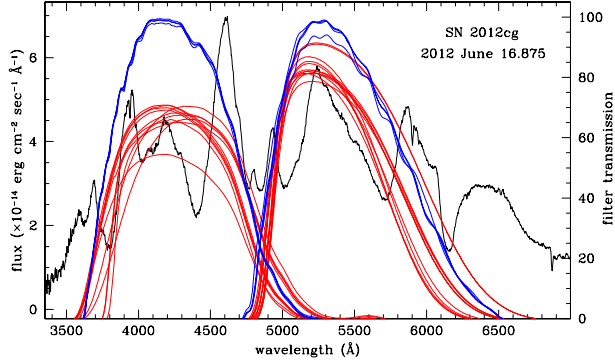}
     \caption{A spectrum of SN 2012cg with overplotted the measured
     transmission profiles of many $B$ and $V$ filters from different
     manufactures (from Munari and Moretti 2012).  Red curves refer to
     common filters made with sandwiches of colored glasses, blue curves to
     multi-layer dielectric filters.  The different amount of transmitted
     absorption and emission features is evident (expecially on the
     red wing of $V$ filter passband), which account for the unavoidable
     differences in the photometry obtained with different telescopes even
     if a given SN is measured against the same identical photometric 
     sequence.}
     \label{fig1}
  \end{figure}  

  \begin{figure}[!Ht]
    \centering   
    \includegraphics[width=7.3cm]{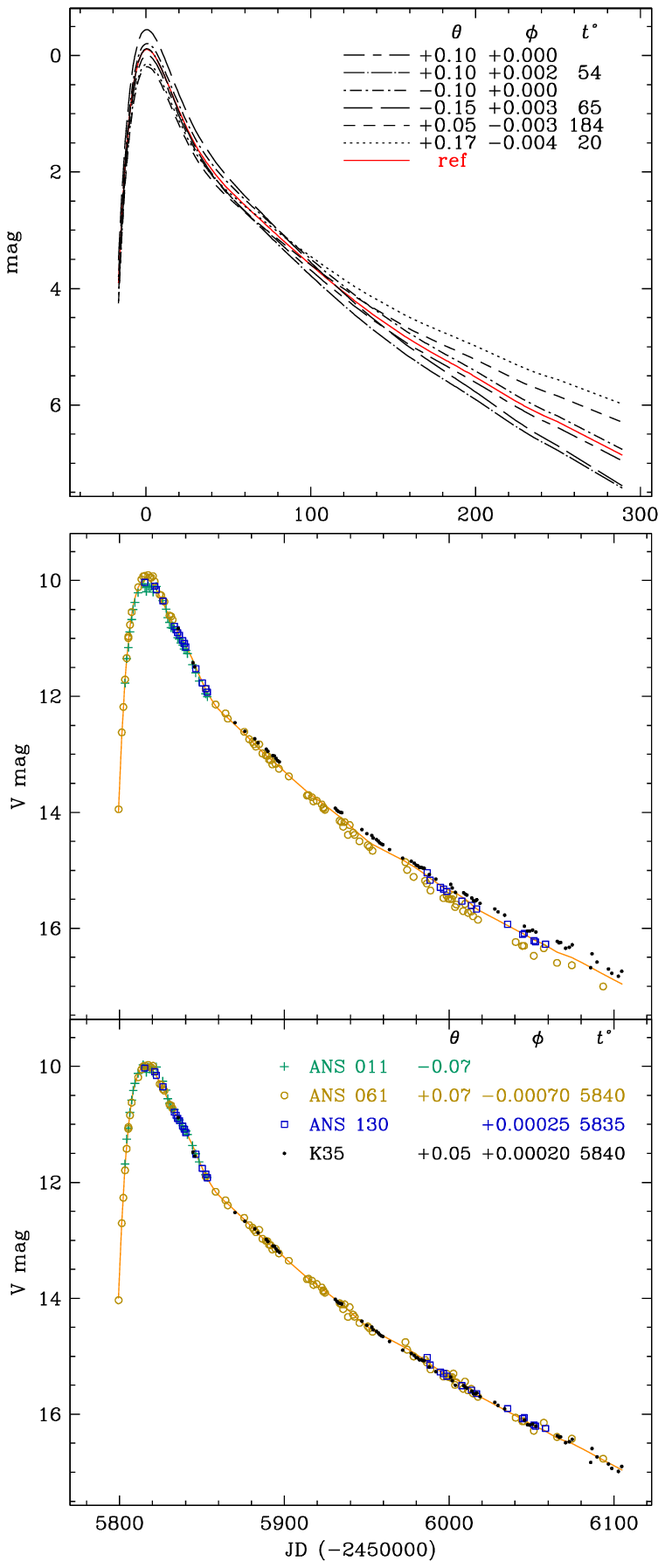}
     \caption{Example of the lightcurve merging method (LMM)
     described in Sect. 3.
     {\it Top:} These lightcurves are the results of shifting 
     (by $\theta$) and stretching (by $\phi$) around the pivot
     points ($t^\circ$) the ordinates (magnitudes) of the reference 
     ligthcurve (plotted in red).  
     {\it Middle:} our source $V$-band observations for SN 2011fe,
     where different colors and symbols identify different telescopes.
     {\it Bottom:} our merged $V$-band lightcurve of SN 2011fe after
     application (by the given quantities) of the LMM shift and pivot 
     stretching method. 
     The orange line (the same in both the middle and bottom panels)
     is a spline fit to the {\it merged lightcurve}
     mag($\lambda$)$_{i,j}$ (cf. Eq.(3)).}
     \label{fig2}
  \end{figure}

\section{Observations}
\label{}

$B$$V$$R_{\rm C}$$I_{\rm C}$ photometry of the three program supernovae was
obtained with several robotic, remotely or manually controlled telescopes 
operated by ANS Collaboration.  Technical details of this
network of telescopes running since 2005, their operational procedures and
sample results are presented by Munari et al.~(2012).  Detailed analysis
of the photometric performances and measurements of the actual transmission
profiles for all the photometric filter sets in use is presented by Munari
and Moretti (2012).  

Additional $B$$V$$R_{\rm C}$$I_{\rm C}$ measurements of SN 2011fe and SN
2012aw were obtained at the Astrokolkhoz Observatory in New Mexico with
K35, a 35cm remotely operated telescope which houses a set of Astrodon
$B$$V$$R_{\rm C}$$I_{\rm C}$ multi-layer dielectric filters.

All measurements were carried out with aperture photometry, the long focal
length of the telescopes and the smoothness of galaxy background around the
three supernovae not requiring the use of PSF-fitting.

All photometric measurements of SN 2011fe and SN 2012aw were carefully tied
to the local $B$$V$$R_{\rm C}$$I_{\rm C}$ sequences calibrated by Henden et
al.  (2012) against Landolt (1983, 1992) equatorial standards.  A similar
sequence was established around SN 2012cg and adopted by all ANS
Collaboration telescopes monitoring this supernova.  The adopted
transformation equations between local realizations (small letters) and the
standard system (capital letter) take the usual form: 

\begin{eqnarray}
        B  &=&  b + \alpha_b \times (b-v) + \gamma_b \nonumber \\              
        V  &=&  v + \alpha_v \times (v-i) + \gamma_v \nonumber \\   
R_{\rm C}  &=&  r + \alpha_r \times (v-i) + \gamma_r \nonumber \\
I_{\rm C}  &=&  i + \alpha_i \times (v-i) + \gamma_i  \\
                B-V  &=&  \beta_{bv} \times (b-v) + \delta_{bv} \nonumber \\
        V-R_{\rm C}  &=&  \beta_{vr} \times (v-r) + \delta_{vr} \nonumber \\
        V-I_{\rm C}  &=&  \beta_{vi} \times (v-i) + \delta_{vi} \nonumber
\end{eqnarray}
where $\alpha$, $\beta$, $\gamma$, $\delta$ are constants describing the
instantaneous realization of the standard system.  Their values, range of
variability, and dependence upon filters, detectors, telescope, and
atmosphere for ANS telescopes are discussed in Munari and Moretti (2012).

The median value of the total error budget (defined as the quadratic sum of
the Poissonian error on the supernova and the formal error on the
transformation from the local to the standard system) for the data collected
on the three supernovae is 0.011 mag for $B$, 0.007 in $V$, 0.010 in $R_{\rm
C}$, 0.016 in $I_{\rm C}$, and 0.010 mag for $B-V$, 0.011 in $V-R_{\rm C}$,
and 0.018 in $V-I_{\rm C}$. Colors and band magnitudes are obtained separately
during the reduction process, and are not derived one from the other.

\section{The lightcurve merging method (LMM)}
\label{}

The spectra of supernovae are dominated by broad emission and absorption
features and are completely different from the black-body like spectra of
normal field stars used in transforming the local photometry to the standard
system.  Local realizations of the photometric system include different
atmospheric and filter transmission, detector sensitivity, efficiency of the
optics.  On normal stars (like those making the photometric comparison
sequence) all these effects are collectively corrected for by the
transformation Eq.  (1), provided that the local realization is sufficiently
close to the standard system (i.e. $\alpha_\lambda$ are close to 0.0, and
$\beta_\lambda$ close to 1.0). The spectra of supernovae, strongly dominated
by broad and intense absorptions and emissions, are completely different
from the black-body like spectra of normal field stars, and the Eq.(1)
transformations can only partially compensate for the differences between
the local and the standard systems.

The situation is illustrated in Figure~1, where the measured transmission
profiles for the $B$ and $V$ band filters in use with ANS Collaboration
telescopes are overplotted with a flux-calibrated spectrum of SN 2012cg which
was obtained on 2012 June 16.875 UT with the Asiago 1.22m telescope of the
University of Padova, operating at 2.31 \AA/pix over the 3300-8100 \AA\
range.  These filters come from different manufacturers and are of both the
multi-layer dielectric type (profiles plotted in blue) and the classical
sandwich of colored glasses (profiles plotted in red), typically built
according to the standard recipe of Bessell (1990).  Some of the profiles shown
in Figure~1 are for brand-new filters, others for filters in long use at the
telescopes and showing clear aging effects (see Munari and Moretti 2012 for
a detailed discussion).

The lightcurves of the same supernova obtained with different telescopes
will therefore show offsets among them, and these offsets will be
time-dependent following the spectral evolution of the supernova.  An
example is shown in Figure~2 (middle panel), where our $V$-band data on SN
2011fe from different telescopes are compared. Which of these lightcurve
is the correct one~? There is no {\em a priori} argument to prefer one or another,
all of them referring to the same photometric calibration sequence, using
the same extraction software, suffering from similar and very low 
formal errors.

The closest representation of the {\it true} lightcurve appears to be a
proper combination of all these individual lightcurves, effectively
averaging over different observing conditions and instrumentation.  To
combine them into a merged lightcurve, we assume that all individual
lightcurves from different telescopes are exactly the same one, affected 
only by an offset in the ordinate zero-point and a linear stretch of the
ordinates around a pivot point:
\begin{eqnarray}
\noindent
{\rm mag_{i,j}(\lambda)} &=& {\rm mag^\ast_{i,j}(\lambda)} + \theta_j(\lambda) 
+ \\ \nonumber
&+& \phi_j(\lambda) \times [t_{i,j}(\lambda) - t_j^\circ(\lambda)]
\end{eqnarray}
where mag$^\ast_{i,j}(\lambda)$ is the $i$-th magnitude value in the
$\lambda$ photometric band from the $j$-th telescope, mag$_{i,j}(\lambda)$
is the corresponding value in the merged lightcurve, $\theta_j(\lambda)$ is
the amount of zero-point shift and $\phi_j(\lambda)$ is coefficient of the
linear stretching around the $t^\circ_j (\lambda)$ pivot point. The stretch
is therefore a linear function of time difference with the reference $t^\circ_j
(\lambda)$ epoch.

The optimal values for $\theta_j(\lambda)$, $\phi_j(\lambda)$ and $t^\circ_j
(\lambda)$ are derived by a $\chi^2$ minimization of the difference between 
the observed points and those defining the merged lightcurve:
\begin{equation}
     \chi^2 = \sum_{i} \sum_{j} \frac{[{\rm mag^\ast_{i,j}(\lambda)} 
              - {\rm mag_{i,j}(\lambda)}]^2}{[\epsilon_{i,j}(\lambda)]^2} 
\end{equation}
where $j$ sums over the different telescopes, $i$ over the individual
measurements of each telescope, and $\epsilon_{i,j}(\lambda)$ is the total
error budget of each measurement. The effect of shifting and stretching
around a pivot point the ordinate values of a given lightcurve is
illustrated in the top panel of Figure~2. The application of the method to
the actual data we obtained for the three supernovae is illustrated in the
middle and bottom panels of Figure~2, using the $V$-band observations of SN
2011fe as an example. The reduction of the dispersion around a common
lightcurve is evident when comparing the middle and bottom panels of
Figure~2.  The coefficients for the other bands, colors, instruments and
supernovae considered in this paper are similarly small.

   \begin{table}[!Ht]
      \caption{Our merged $B$$V$$R_{\rm C}$$I_{\rm C}$ photometry for SN
      2011fe (the table is published in its entirety in the
      electronic edition of New Astronomy.  A portion is shown here for
      guidance regarding its form and content).}
      \centering
      \includegraphics[width=7.8cm]{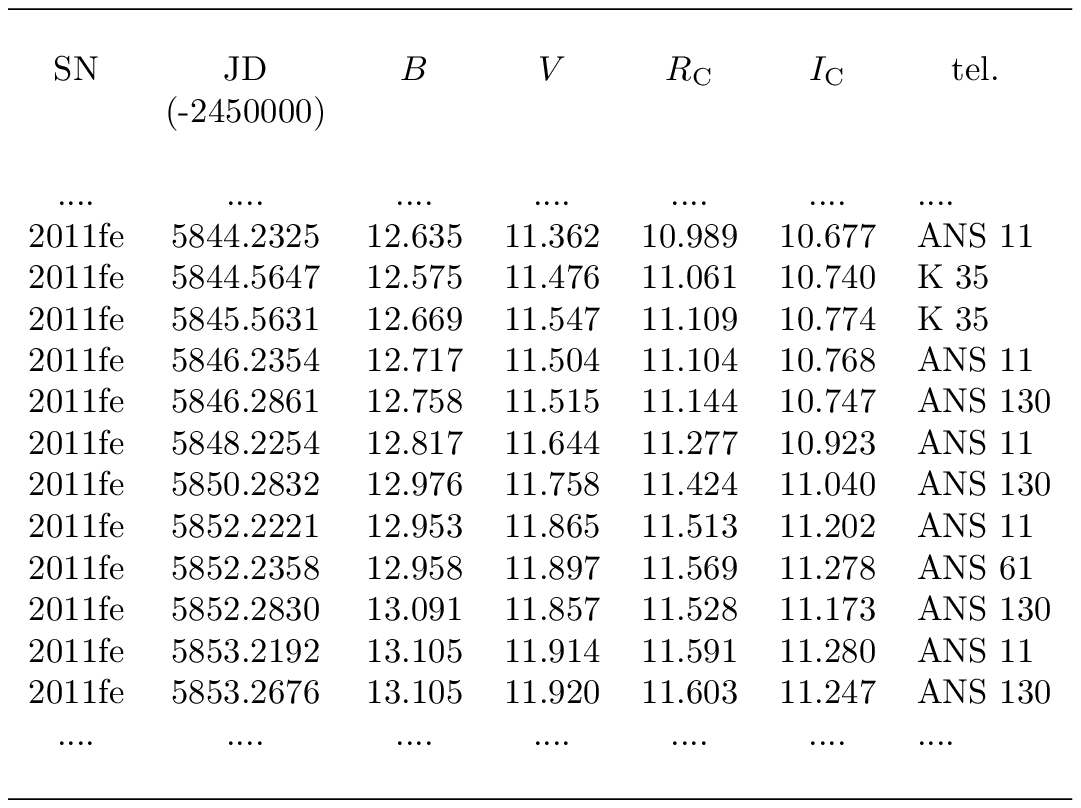}
      \label{tab1}
  \end{table}

   \begin{table}[!Ht]
      \caption{Our merged $B$$V$$R_{\rm C}$$I_{\rm C}$ photometry for SN 2012cg.}
      \centering
      \includegraphics[width=7.8cm]{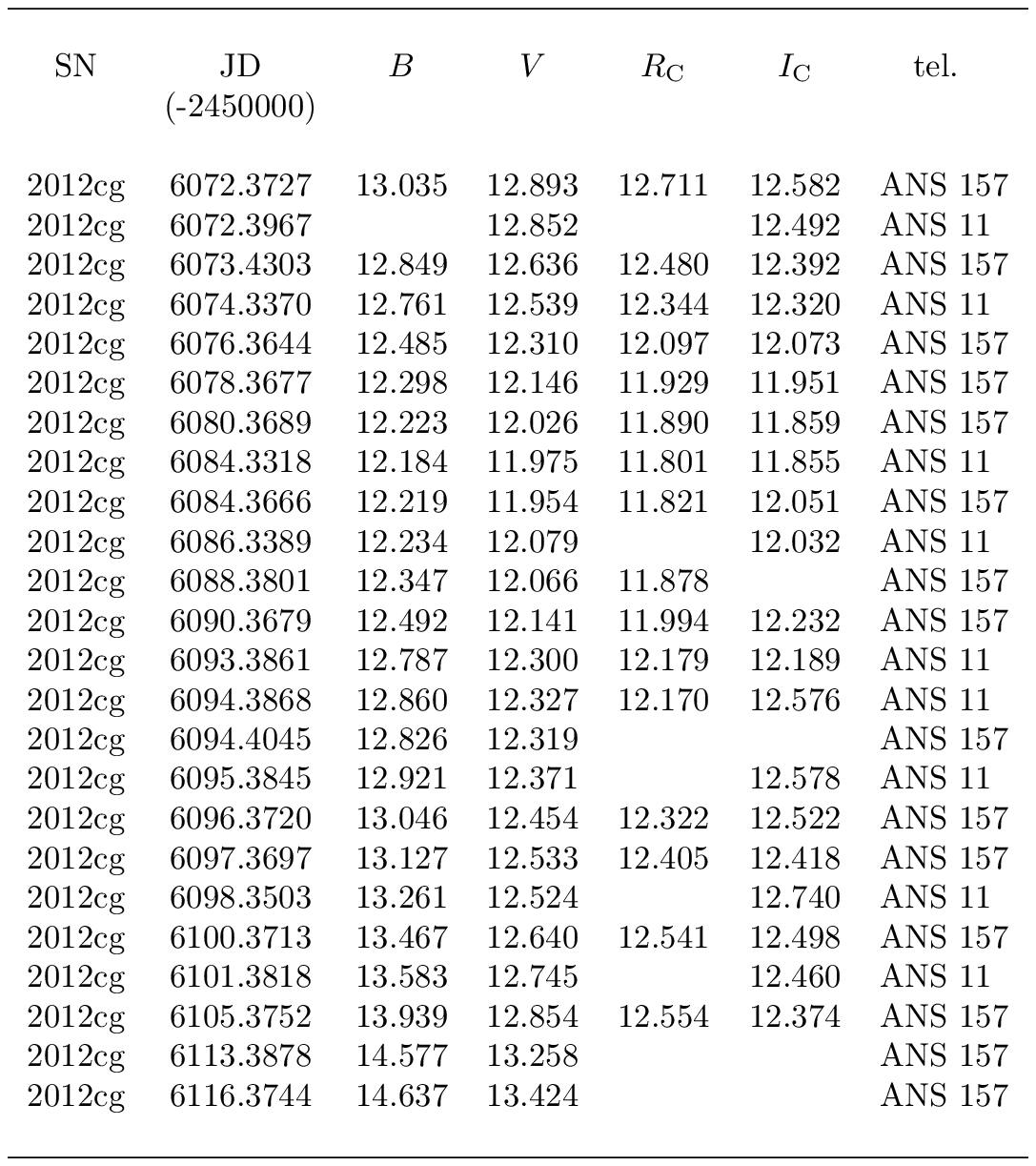}
      \label{tab2}
  \end{table}

   \begin{table}[!Ht]
      \caption{Our merged $B$$V$$R_{\rm C}$$I_{\rm C}$ photometry for SN
      2012aw (the table is published in its entirety in the
      electronic edition of New Astronomy.  A portion is shown here for
      guidance regarding its form and content).}
      \centering
      \includegraphics[width=7.8cm]{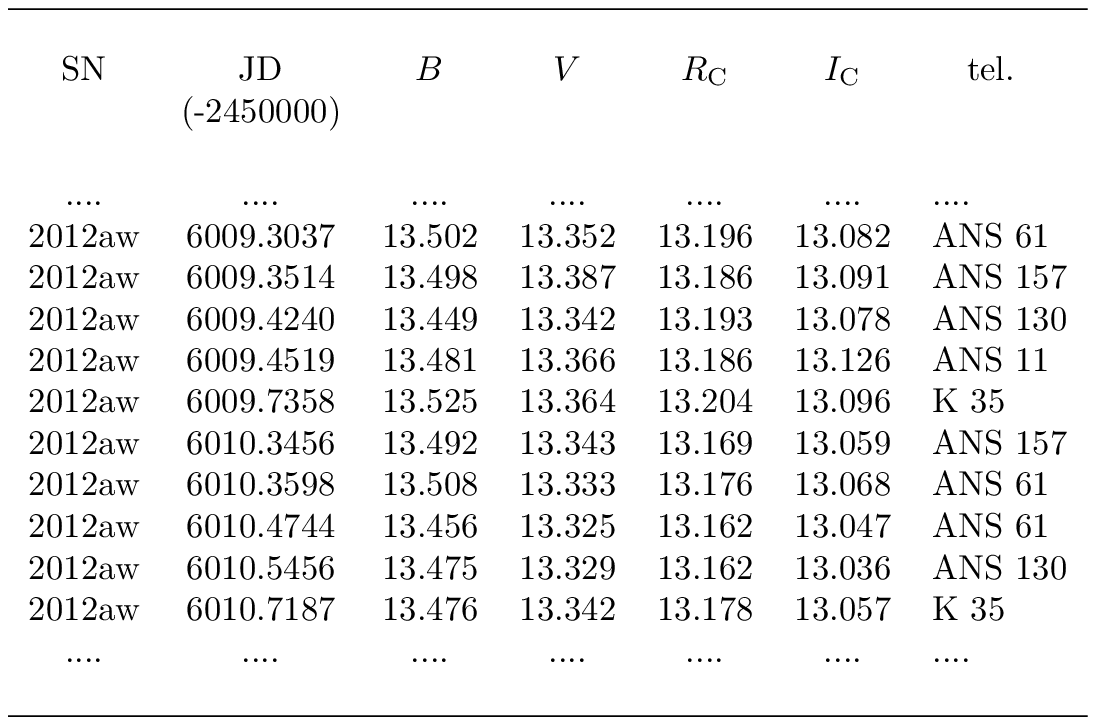}
      \label{tab3}
  \end{table}

Notice that the LMM method to construct a merged
lightcurve from observations obtained with different telescopes {\it does
not stretch} the resulting lightcurve in flux (compare how
the orange line, the same curve in both panels, goes through the data in the
middle and bottom panels of Figure~2), nor time (the
method acts only on the individual magnitude values, not the associated time
tag). Therefore, the LMM method does not interfere with the application of
popular time- and brightness-stretching methods used to estimate the
absolute magnitude of a supernova by comparison with a sample of template
ligthcurves (see below).

  \begin{figure*}[!t] \centering
    \includegraphics[width=16.0cm]{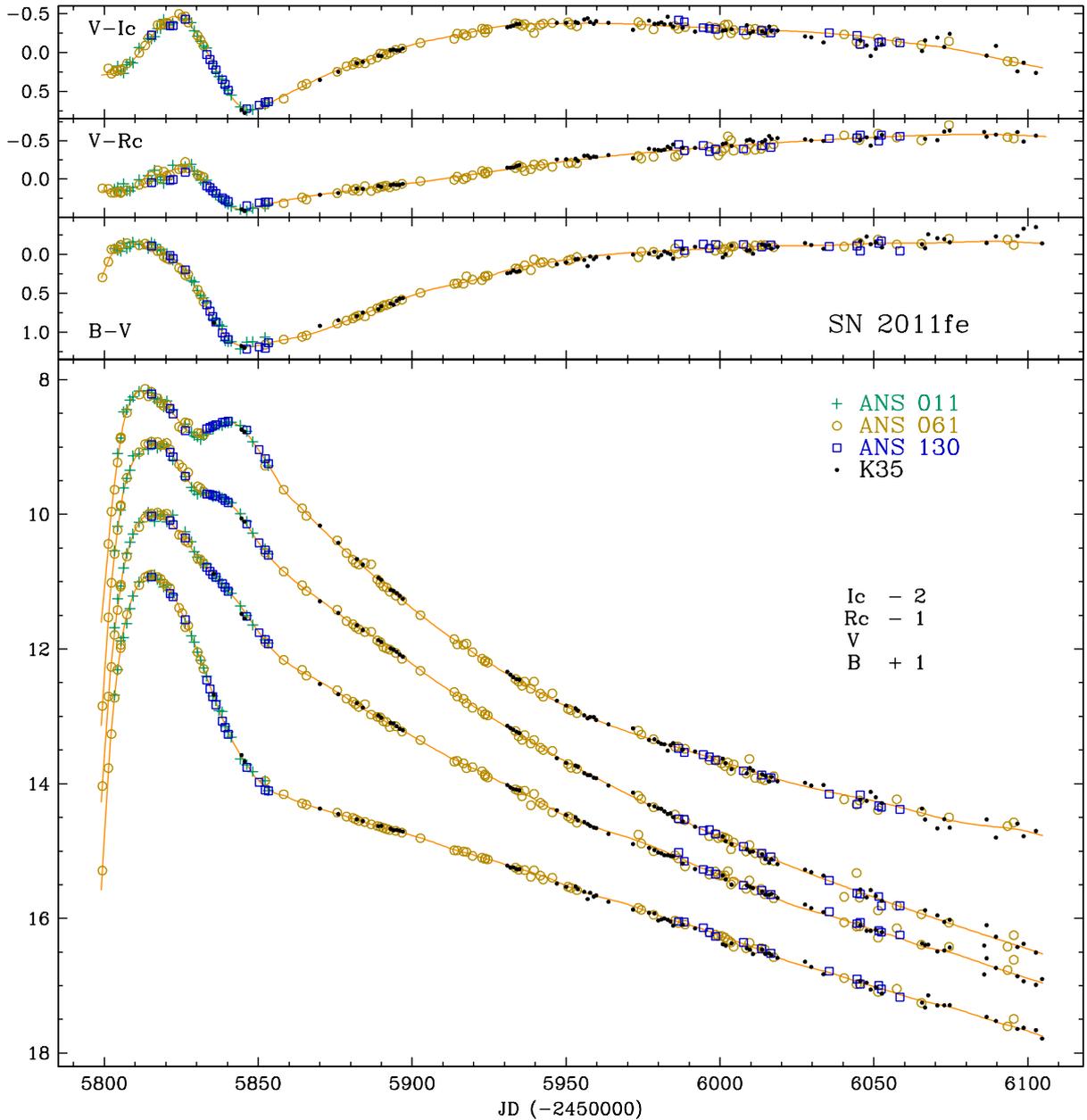}
     \caption{$B$$V$$R_{\rm C}$$I_{\rm C}$ light- and color-curve of SN
     2011fe from our data in Table~1.  Different colors and symbols identify
     the telescopes used to monitor the supernova. The curves are spline 
     fits to the data drawn to guide the eye.}
     \label{fig3} 
   \end{figure*}

  \begin{figure*}
    \centering   
    \includegraphics[height=14.2cm,angle=270]{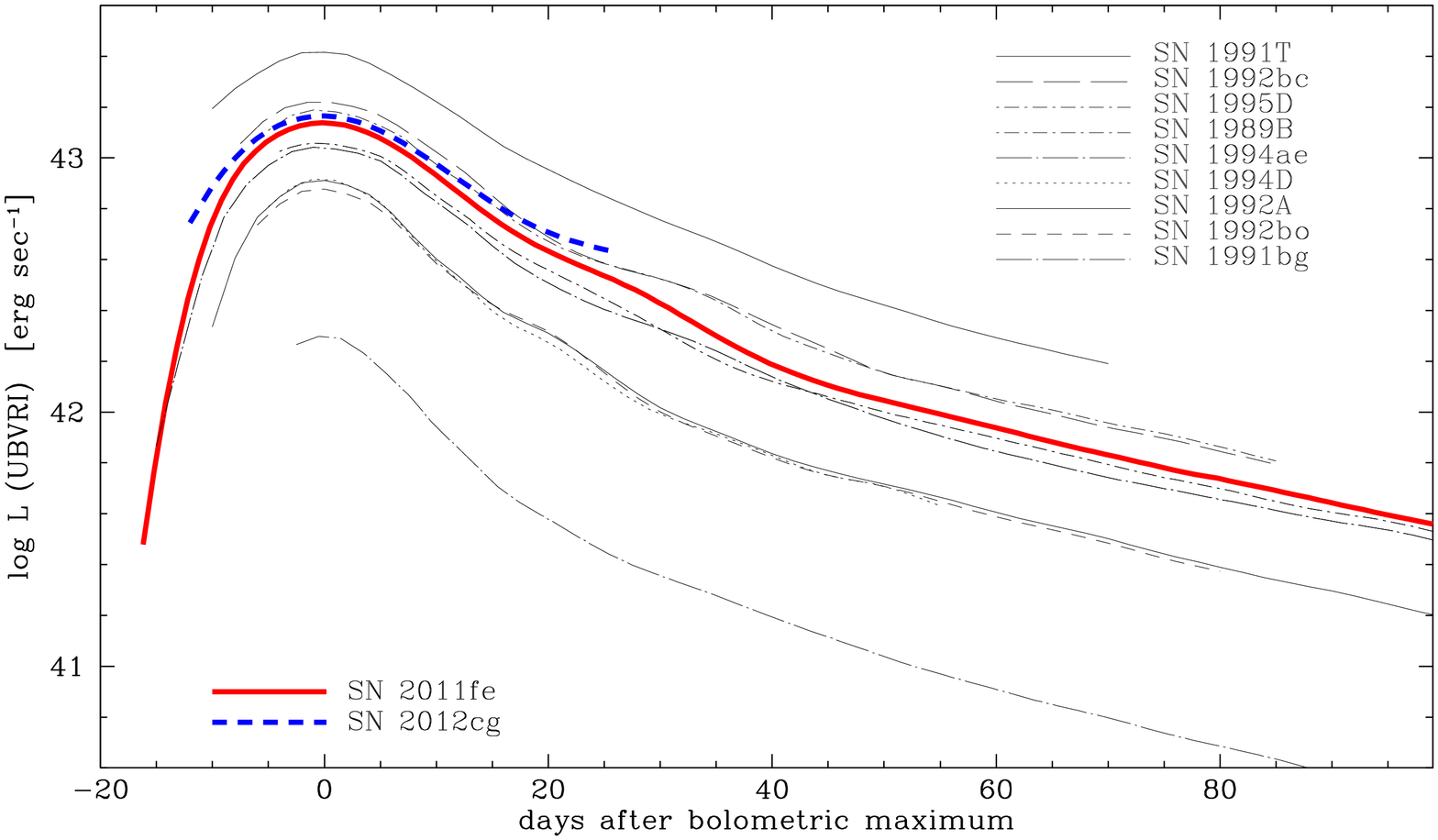}
     \caption{Bolometric lightcurves of SN 2011fe and SN 2012cg (obtained by summing the
              reddening corrected flux in the $U$$B$$V$$R_{\rm C}$$I_{\rm
              C}$ bands) compared to those of well observed objects from
              Contardo et al.  (2000).  The $U$ data are obtained from
              observed $B$ values and adding the parametrized $U$$-$$B$
              color evolution from Nobili and Googar (2008).}
     \label{fig4}
  \end{figure*}  

Our merged $B$$V$$R_{\rm C}$$I_{\rm C}$photometry for the three program
supernovae is listed in Tables~1,2 and 3, and presented in the Figures 3, 5,
6 and 7.  All tegether, we present 236 $B$$V$$R_{\rm C}$$I_{\rm C}$ data
sets for SN 2011fe, 108 for SN 2012aw, and 27 for SN 2012cg.

\section{Results}
\label{}

\subsection{SN 2011fe}

The type Ia SN 2011fe was discovered in M101 at $g$=17.35 on 2011 Aug 24.164
UT by the Palomar Transient Factory (Nugent et al.~2011a), only $\sim$11
hours after it exploded (on Aug 23.71, Nugent et al.~2011b).  From archival
HST images, Li et al.~(2011) excluded a luminous red giant as a companion
star to the progenitor of SN 2011fe, a conclusion supported by sensitive
X-ray and radio non detection during early evolution (Horesh et al.  2012,
Chomiuk et al.  2012).  Analysis of early-time optical spectra was reported
by Parrent et al.  (2012), ultraviolet data from Swift satellite by Brown et
al.  (2012), and polarization of optical light by Smith et al.  (2012). 
Being the first close type Ia supernova detected in the CCD era, SN 2011fe
will undoubtedly become the best observed thermonuclear supernova, well into
its nebular stage, and a stringent test for theoretical models.

We began our photometric monitoring of SN 2011fe on 2011 August 25.891, less
than a day after the announcement of discovery was posted by Nugent et al. 
(2011a).  We have obtained 236 independent $B$$V$$R_{\rm C}$$I_{\rm C}$
runs, the last on 2012 August 1.404 UT, covering 342 days.  We began when
the SN was still 4.4 mag below maximum, and continued until it declined by
7.2 mag below it.  Table~4 lists the time and brightness of the optical
maximum from our observations, both as observed and as corrected for the
$E_{B-V}$=0.025$\pm$0.003 reddening found by Patat et al.  (2011) to affect
SN 2011fe.  In correcting for the reddening we have adopted a standard
$R_V$=3.1 reddening law (Fitzpatrick 1999) and the reddening relations of
Fiorucci and Munari (2003) appropriate for the observed colors.

   \begin{table}[!t]
      \caption{Time and brightness of SN 2011fe at maximum. The fourth column
      gives the decline in magnitude 15 days past maximum. The last two
      columns list the brightness at maximum and the distance modulus 
      corrected for $E_{B-V}$=0.025 reddening, adopting the standard 
      $R_V$=3.1 exctinction law and Prieto et al. (2006) calibrations of 
      $\Delta$mag$_{15}$.}
      \centering
      \includegraphics[width=6.8cm]{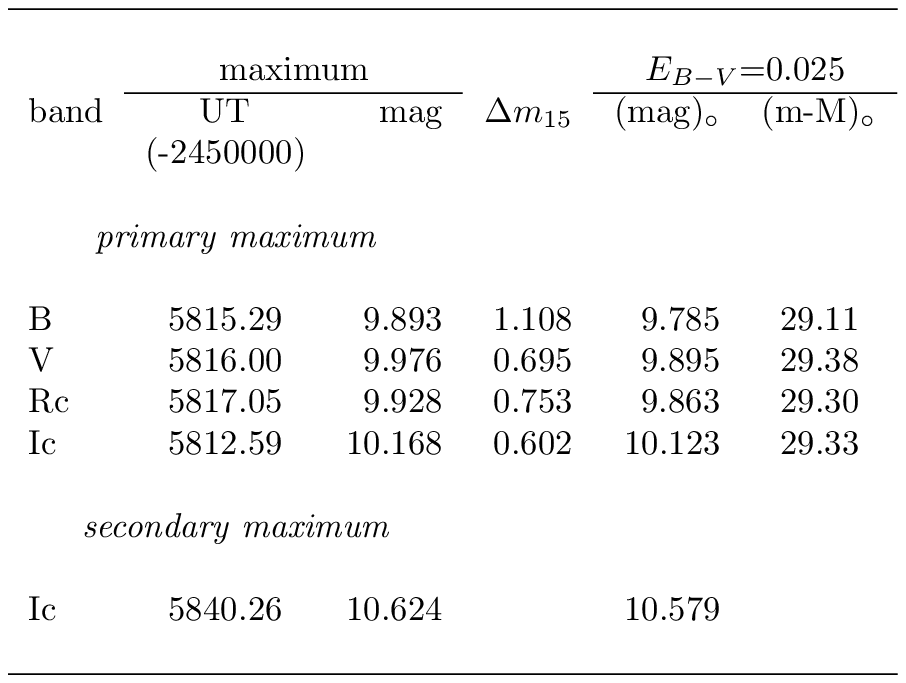}
      \label{tab4}
  \end{table}

Our epochs and magnitudes at maximum brightness are somewhat different from
those listed by Richmond and Smith (2012), who presented $B$$V$$R_{\rm
C}$$I_{\rm C}$ photometry covering the first 180 days of SN 2011fe
evolution.  The differences are probably due to the preliminary AAVSO
comparison sequence that they used, based on only three stars, of similar
colors and much redder ($B$$-$$V$=+0.63, +0.68, +0.84) than the supernova at
maximum brightness ($B$$-$$V$=$-$0.08).  This forced Richmond and Smith to
adopt the values of $\alpha_b$, $\alpha_v$, $\alpha_r$, $\alpha_i$ in the
transformation equations Eq.(1) above from unrelated observations of a
single Landolt (1992) field, and use the three AAVSO stars around SN 2011fe
only to constrain the zeropoints $\gamma_b$, $\gamma_v$, $\gamma_r$,
$\gamma_i$ in the same equations.  Appreciably closer to our Table~4 are the
values obtained by Tammann and Reindl (2011) from analysis of photometry
collected by AAVSO members, covering the first 75 days of SN 2011fe
lightcurve.

The absolute magnitude of type Ia supernovae is related to the shape of
their lightcurves, with Phillips (1993) noting how the intrinsic brightness
is inversely proportional to the speed of decline.  The most used
calibrations of the relation between absolute brightness and rate of decline
are probably the $\Delta m_{\rm 15}$, MLCS and stretch methods. 
The $\Delta m_{\rm 15}$ method is based on the magnitude difference between
maximum light and 15 days past it (Hamuy et al.  1996, Phillips et al. 
1999).  The MLCS method (Riess et al.~1996, 1998)
fits the observed lightcurve to a set of template lightcurves
built from a set of well observed and calibrated supernovae.  The stretch
method (Permutter et al.  1997; Goldhaber et al.  2001) derives the decline
speed and therefore the absolute magnitude by stretching the scale on the
time axis of the lightcurve to fit a reference lightcurve.  
Prieto et al. (2006) presented a new technique that is a combination of the 
$\Delta m_{\rm 15}$ and MLCS methods. Their calibration of the absolute magnitude 
at maximum takes the form:
\begin{equation}
M_\lambda (max) = a_\lambda + b_\lambda(\Delta m_{\rm 15} - 1.1)
\end{equation}
where $\lambda$=$B$,$V$,$R_{\rm C}$,$I_{\rm C}$.  Our determination of 
$\Delta m_{\rm 15}$ in the $B$,$V$,$R_{\rm C}$,$I_{\rm C}$ bands is 
given in Table~4, together with the resulting distances to SN 2011fe.

The distance derived from $\Delta m_{\rm 15}$ in the $V$,$R_{\rm C}$,$I_{\rm
C}$ bands are tightly grouped around the mean value
$<(m-M)_\circ>$=29.34$\pm$0.02, which is very close to the
$(m-M)_\circ$=29.39$\pm$0.04 distance to M101 derived by Tammann and Reindl
(2011) from HST observations of stars at the tip of the RGB by Sakai et al. 
(2004) and Rizzi et al.  (2007).  The distance obtained from $\Delta
m_{\rm 15}$ in the $B$ band is significantly less, by 0.23 mag,
but close to the distance $(m-M)_\circ$=29.05$\pm$0.13 that Shappee and
Stanek (2011) derived for M101 from HST observations of Cepheid variables. 
We will see in the next paragraph, that a similar situation exists
with SN 2012cg, where the Prieto et al.  (2006) relations for $V$,$R_{\rm
C}$,$I_{\rm C}$ provide comparable distances, while in the $B$ band
the distance is less by 0.22 mag.  This suggests a possible revision of
zero-points of Prieto et al.  (2006) for the $B$ band:
\begin{equation}
M_B {\rm (max)} = -19.550 + 0.636 \times (\Delta B_{\rm 15} - 1.1)
\end{equation}
for the $E_{B-V}\leq0.06$ low reddening case, and
\begin{equation}
M_B {\rm (max)} = -19.511 + 0.753 \times (\Delta B_{\rm 15} - 1.1)
\end{equation}
for the $E_{B-V}>$0.06 case. At least for SN 2011fe and SN 2012cg, adding 
$-$0.225 mag to the zero-points would bring the distance derived from $B$ band
into agreement with those obtained from the $V$,$R_{\rm C}$,$I_{\rm C}$ bands. 
It is worth noticing that both these supernovae are close to $\Delta B_{\rm
15}$=1.1, so changing the slope instead of the zero-points would not cure the
systematic shift between the distance in $B$ and that derived in the $V$,$R_{\rm
C}$,$I_{\rm C}$ bands. The absolute magnitude of SN 2011fe corresponding to
the $(m-M)_\circ$=29.34 distance modulus is $M_B$=$-$19.55. The secondary
maximum in the $I_{\rm C}$ band was fainter by 0.46 mag and occurred 28.0 days 
later, both values being well within the range observed for
type Ia supernovae (eg. Leibundgut 2000).

  \begin{figure}[!t]
    \centering   
    \includegraphics[width=8cm]{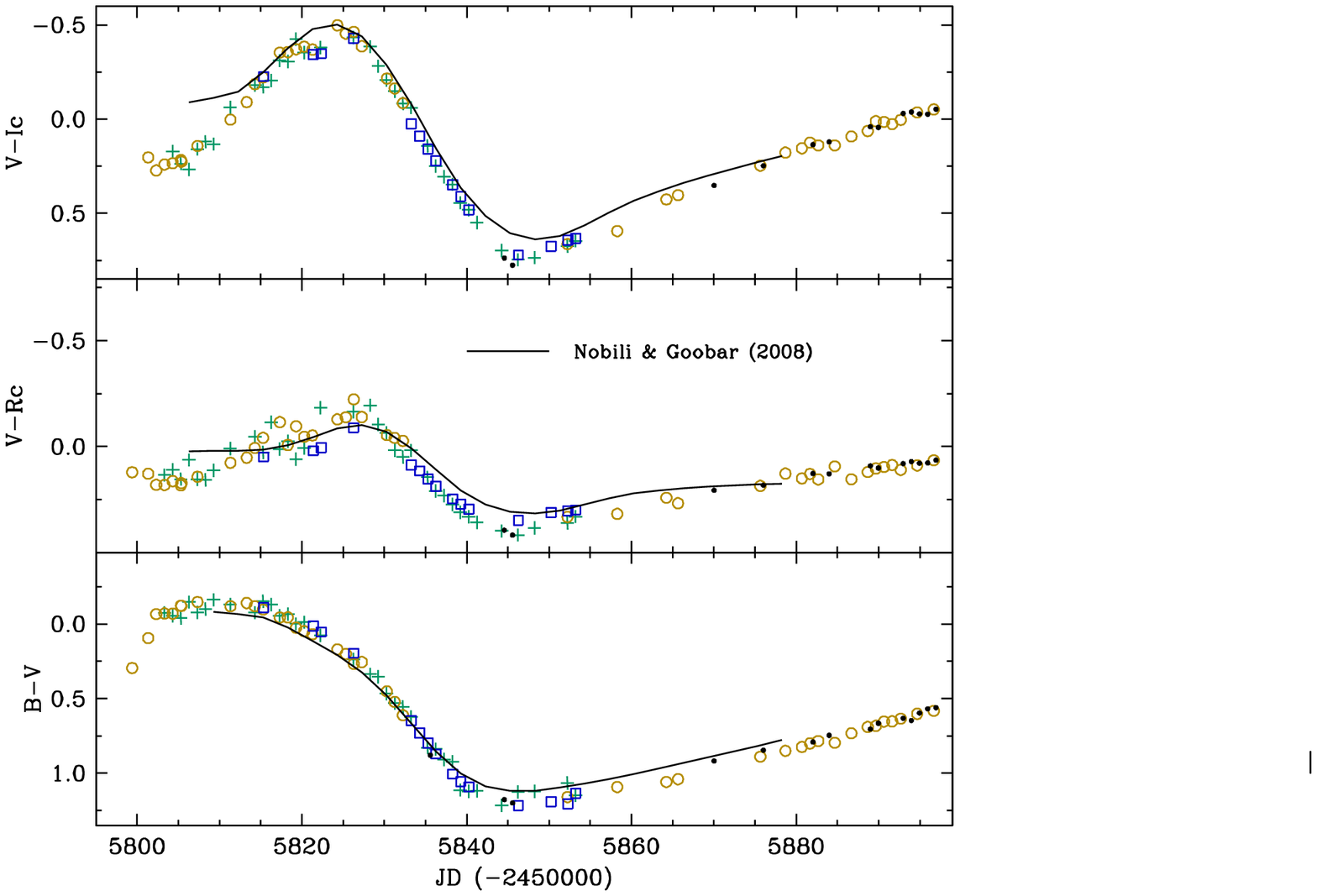}
     \caption{Comparison of the color evolution of SN 2011fe with
     the parameterized color evolution of Nobili and Goobar (2008),
     over-plotted as a curve.}
     \label{fig5}
  \end{figure}  

The bolometric curve of SN 2011fe is presented in Figure 4, where it is
compared to the bolometric evolution of a sample of well observed type Ia
supernovae compiled by Contardo et al.  (2000).  All the bolometric
lightcurves in Figure~4 are obtained by adding the flux observed in the
$U$$B$$V$$R_{\rm C}$$I_{\rm C}$ bands, which is a close approximation of the
{\em true} bolometric lightcurve (obtained integrating the flux over the
whole wavelength range), given the evidence that about 80\% of the
bolometric luminosity of a typical SN Ia is emitted over the optical range
(Suntzeff 1996).  We did not observe SN 2011fe in the $U$ band, but obtained
it from our $B$-band data and the $U$$-$$B$ average color evolution of type
Ia supernovae as given by Nobili and Goobar (2008).  This is a safe
procedure because $U$ band is a marginal contributor to the optical flux, and
SN 2011fe followed the typical $B$$-$$V$, $V$$-$$R_{\rm C}$ and
$V$$-$$I_{\rm C}$ color evolution of type Ia supernovae (see Figure~5).  The
shape of the bolometric lightcurve nicely follows the progression with
absolute magnitude presented by Contardo et al.~(2000).  The maximum of the
bolometric luminosity is reached $\sim$18.2 days past the beginning of eruption
as determined by Nugent et al.  (2011b), close to the 18.03$\pm$0.24 days found
by Ganeshalingam et al.~(2011) as the mean value for spectroscopically
normal type Ia supernovae. 

\subsection{SN 2012cg}

SN 2012cg was discovered in NGC 4424 on 2012 May 17.22 UT by Kandrashoff et
al.~(2012) during the Lick Observatory Supernova Search (LOSS), and soon
spectroscopically identified as a type Ia supernova caught well before
optical maximum.  An X-ray observation with Swift obtained 1.5 days after
the discovery failed to detect emission from the supernova (Margutti and
Soderberg 2012).  Post-discovery inspection of MASTER-Kislovodsk images of
NGC 4424 obtained on May 15.790, reveals the supernova was at that time
already present at a magnitude $\sim$19 (Lipunov 2012).

We began our observations of SN 2012cg on May 24.873 UT, 9 days before $B$
maximum, and continued them for 44 days.  We obtained a total of 27
$B$$V$$R_{\rm C}$$I_{\rm C}$ measurement data sets.

   \begin{table}[!Ht]
      \caption{Time and brightness of SN 2012cg maximum. The fourth column
      gives the decline in magnitude 15 days past maximum. The last two
      columns list the brightness at maximum and the distance modulus 
      corrected for $E_{B-V}$=0.18 reddening, adopting the standard      
      $R_V$=3.1 exctinction law and Prieto et al. (2006) calibrations of  
      $\Delta$mag$_{15}$.}
      \centering
      \includegraphics[width=6.8cm]{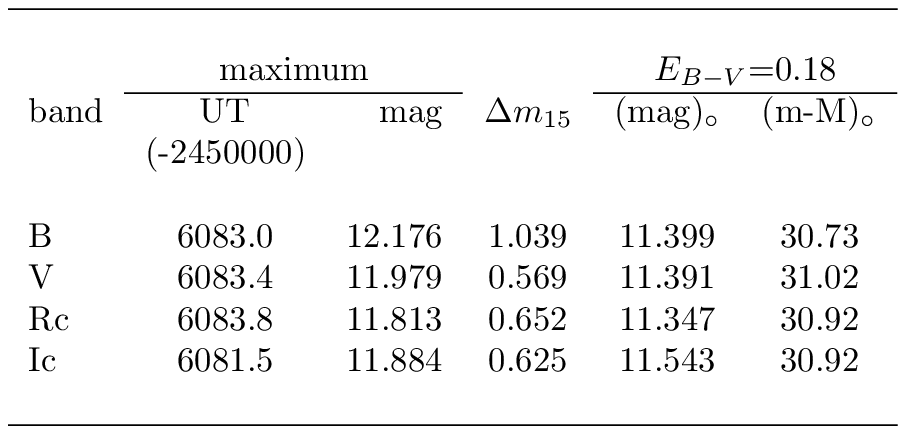}
      \label{tab5}
  \end{table}

  \begin{figure}[!Ht]
    \centering   
    \includegraphics[width=8cm]{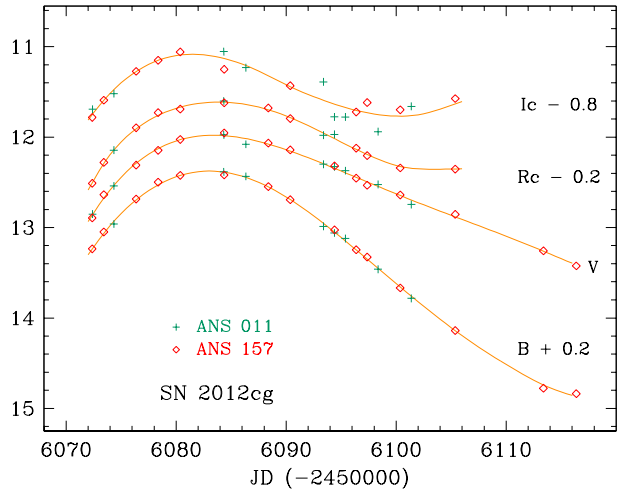}
     \caption{$B$$V$$R_{\rm C}$$I_{\rm C}$ lightcurve of SN
     2012cg from our data in Table~2.  Different colors and symbols identify
     the telescopes used to monitor the supernova. The curves are spline 
     fits to the data drawn to guide the eye.}
     \label{fig6}
  \end{figure}  

The time and brightness of SN 2012cg at maximum are given in Table~5. The
$B$-band maximum corresponds to June 4.5 UT.  Marion et al.  (2012)
similarly found the maximum to have occurred around June 4.  The indication
by Silverman et al.  (2012) of a $B$-band maximum occurring on June
2.0$\pm$0.75 cannot be easily reconciled with our observations. 

  \begin{figure*}[!t]
    \centering   
    \includegraphics[width=16.0cm]{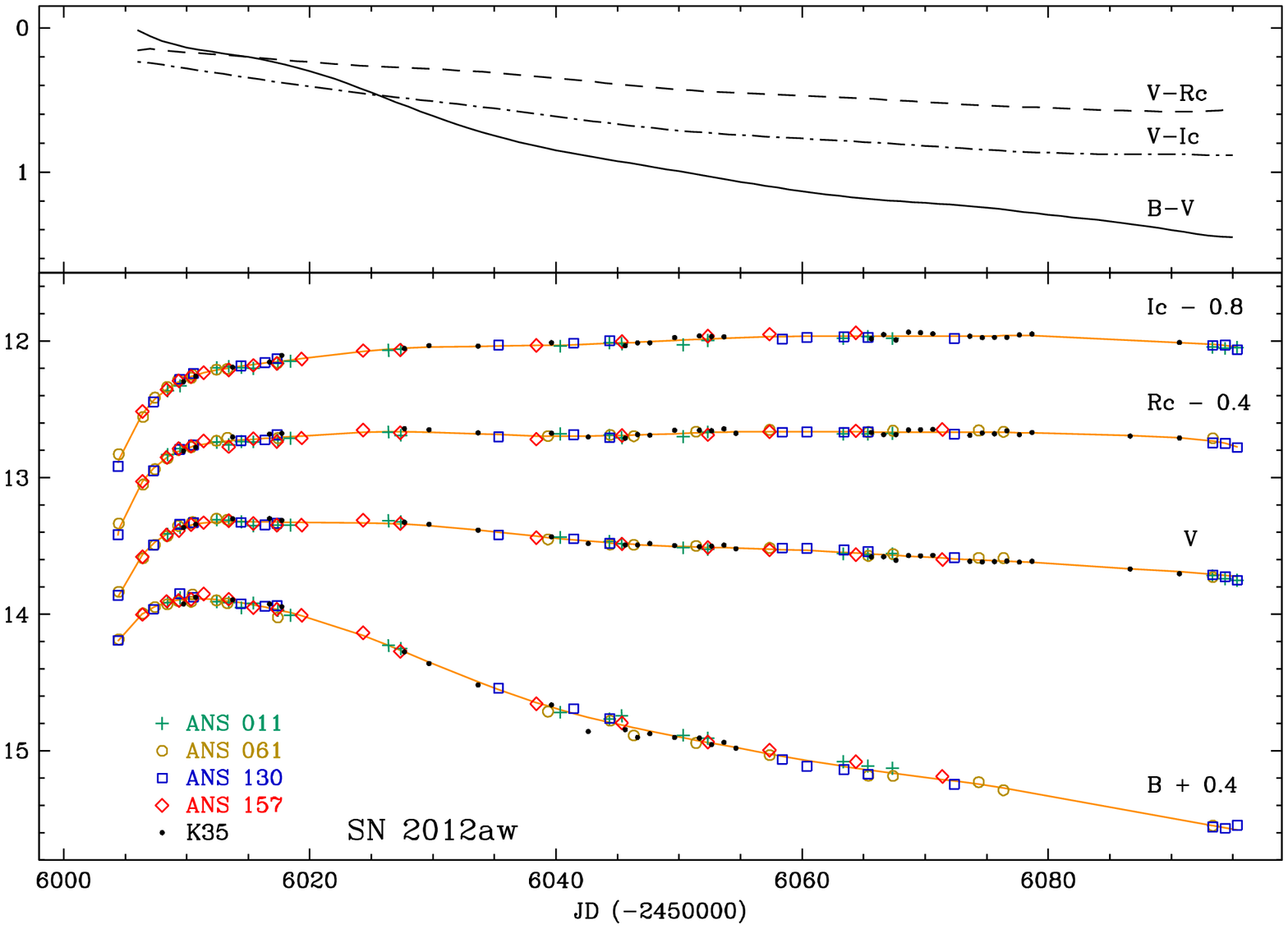}
     \caption{$B$$V$$R_{\rm C}$$I_{\rm C}$ light- and color-curve of
     SN 2012aw from our data in Table~3.  Different colors and symbols identify
     the telescopes used to monitor the supernova. The curves are spline 
     fits to the data drawn to guide the eye.}
     \label{fig7}
  \end{figure*}  

The $\Delta m_{\rm 15}$ decline rates of SN 2012cg in Table~5 are somewhat
slower than those of SN 2011fe, suggesting an intrinsically brighter target. 
By correcting for a $E_{B-V}$=0.18 reddening (as suggested by Marion et al. 
2012, and Silverman et al.~2012), we obtain from $V$$R_{\rm C}$$I_{\rm C}$
a distance modulus of (m$-$M)$_\circ$=30.95.  This is quite close to the
(m$-$M)$_\circ$=30.91 obtained for NGC 4424 from the Tully-Fisher relation
by Cortes et al.  (2008).  The corresponding absolute magnitude for SN
2012cg would be $M_B$=$-$19.55, the same as found above for SN 2011fe, in
spite of the slower decline rates.  The bolometric lightcurve of SN 2012cg
plotted in Figure~4 peaks at a luminosity close to that of SN 2011fe on June
4.75, but the rising and declining time are slower than those of SN 2011fe. 
Silverman et al.  (2012) modeled the early rise in magnitude of SN 2012cg
following the usual expanding fireball approach, and by extrapolation
estimated that the supernova explosion occoured on May 15.7$\sim$0.02.  Our
bolometric lightcurve reaches its maximum 20.0 days later.

\subsection{SN 2012aw}

SN 2012aw was discovered by P. Fagotti on Mar 16.86 UT on images of M95
taken at his private observatory when it was at $R_{\rm C}$$\sim$15 (CBET
3054), and was classified from spectra as a type IIP supernova by Itoh et
al.  (2012) and Siviero et al.  (2012).  We began our observations on Mar
17.91 UT, a week before maximum was reached in the $B$ band, and continued
them for 91 days, collecting 108 $B$$V$$R_{\rm C}$$I_{\rm C}$ measurement
data sets.

Freedman et al. (2001) obtained from observation of Cepheids a distance to
the parent M95 galaxy of (m-M)$_\circ$=30.0$\sim$0.1.  While the Milky Way
reddening toward M95 is low ($E_{B-V}$=0.028, Schegel et al.~1998), the
amount local to SN 2012aw seems to be relevant.  Fraser et al.~(2012)
identified on archival HST and ground-based images the supernova progenitor
as a star of $V$$\sim$26.7, appearing red in color and suffering from a
large extinction ($E_{B-V}$$>$0.8), whereas the SN itself does not appear to
be significantly extinguished, a fact that Fraser et al.  interpret as
evidence for the destruction of pre-existing, circumstellar dust during the SN
explosion.  This conclusion is shared by van Dyk et al.  (2012) that found
the progenitor to be a red-supergiant of $M_{\rm bol}$=$-$8.3, being
extinguished by $A_V$$\sim$3.1 mag.

Our photometry shows a long lasting, flat maximum in $V$$R_{\rm C}$$I_{\rm
C}$ bands, distinctive of type IIP supernovae, which is expected to
last about a hundred days (Doggett and Branch 1985), i.e.  longer than our
monitoring period.  The maximum in the $B$ band was reached on Mar 24.1 UT
at $B$=13.487, providing an absolute magnitude of $M_B$=$-$16.4 (correcting
for the Milky Way reddening but not for any residual reddening from dust
local to the SN).  The $\beta^B_{\rm 100}$ decline parameter, as defined by
Patat et al.  (1994), is 2.67, right in the middle of their distribution for
type IIP supernovae, and $\beta^V_{\rm 100}$ is 0.63.  The rate of decline
at the center of the decline branch before the inflection point was $\Delta
B$=0.035 mag day$^{-1}$, and $\Delta B$=0.015 mag day$^{-1}$ at the center
of the plateau phase.  At the time of $B$ maximum, the observed colors of SN
2012aw were $B$$-$$V$=+0.146, $V$$-$$R_{\rm C}$=+0.174, $V$$-$$I_{\rm
C}$=+0.290. They monotonically evolved toward redder colors during the time
covered by our observations, as typical for type IIP supernovae. 

The maximum in $V$ band was reached at $V$=13.321 on Mar 27.2 UT, and the
mean decline rate from then to the end of our monitoring (82 days later) is
0.0050 mag day$^{-1}$.  Over the period of our observations in Figure~7, the
$R_{\rm C}$ lightcurve is best describe as {\em flat} (within $\pm$0.02 mag
of ) once the initial rise was completed, and the $I_{\rm C}$ as showing a 
minimal curvature indicative of a ill-defined, broad maximum around May 24,
two months past the $B$-band maximum. A similar delay in the $I_{\rm C}$
maximum and the flat appearance of the $R_{\rm C}$ lightcurve during the
initial hundred days was observed by Elmhamdi et al.~(2003) for supernova
type IIP 1999em in NGC 1637.

Our lightcurve of SN 2012aw does not show a brightness {\it spike} around
$B$-band maximum as reported by Elmhamdi et al.~(2003) for the type IIP
supernova 1999em brightness, and by Ruitz-Lapuente et al.  (1990) for type
IIP supernova 1988a.  These spikes in the earliest portion of the recorded
lightcurve were attributed to the supernova ejecta slamming onto
circumstellar material originated by mass loss from the progenitor during
the immediate pre-supernova phase.  The reality of these spikes is however
dubious, given the discussion in sect.3 about the differences between local
realizations of a common photometric system.  The lightcurve of 1999em
around maximum brightness was compiled by Elmhamdi et al.  assembling few
data points from different telescopes, mostly contributing a single
observation.  Similarly, the lightcurve of SN 1988a was compiled by
Ruitz-Lapuente et al.  from visual estimates by different observers, each
one contributing only sparse data, and the reality of the supposed spike
rests entirely on only two visual estimates, that come from a single
observer that did not provide other data on this supernova and whose
consistency with the other observers therefore cannot be tested.

\section{Acknowledgments}

We would like to thank Roberto Barbon for a careful reading of an advanced
draft of the present paper, and Andrea Frigo for his continuing support to
the data reduction process within ANS Collaboration.

 \end{document}